\documentclass[graybox]{svmult}

\usepackage{mathptmx}       
\usepackage{helvet}         
\usepackage{courier}        
\usepackage{makeidx}         
\usepackage{graphicx}        
\usepackage{multicol}        
\usepackage[bottom]{footmisc}
\usepackage[utf8]{inputenc}
\usepackage{amsmath}

\makeindex             

\def\non{\nonumber}
\def\br{\begin{eqnarray}}
\def\er{\end{eqnarray}}

\def\({\left(}
\def\){\right)}
\def\[{\left[}
\def\]{\right]}

\def\a{\alpha}

\def\ph{\phi}

\def\pp{\partial}

\def\s{\sigma}

\def\non{\nonumber}
\def\ba{\begin{align}}
\def\ea{\end{align}}
\def\be{\begin{eqnarray}}
\def\ee{\end{eqnarray}}

\def\a{\alpha}

\def\s{\sigma}

\def\ph{\phi}

\def\pp{\partial}

\begin{document}

\title*{The sinh-Gordon defect matrix generalized for $n$ defects}

\author{N.I. Spano, A.L. Retore, J.F. Gomes,  A.R. Aguirre and A.H. Zimerman }
\institute{A.R. Aguirre \at Instituto de F\'isica e Qu\'imica, Universidade Federal de Itajub\'a - IFQ/UNIFEI, Av. BPS 1303,  37500-903, Itajub\'a, MG, Brasil. \email{alexis.roaaguirre@unifei.edu.br}
\and J.F. Gomes, A.L. Retore, N.I. Spano, and A.H. Zimerman  \at Instituto de F\'isica Te\'orica - IFT/UNESP, Rua Dr. Bento Teobaldo Ferraz 271, Bloco II, 01140-070, S\~ao Paulo, Brasil. \email{jfg@ift.unesp.br, retore@ift.unesp.br, natyspano@unesp.br, zimerman@ift.unesp.br}}

\maketitle

\abstract{In this paper we obtain a general expression for 
the n-defect matrix  for the sinh-Gordon model.   This in turn generate 
  the general Bäcklund transformations (BT) for a system with $n$ type-I defects, through a gauge transformation.}

\section{Introduction}
\label{intro}

Integrable models are  known to  be characterized by   an infinite number of conservation laws which are responsible  for the stability of soliton solutions. 
In fact,  these  conservation laws may be regarded as hamiltonians generating time evolutions  within a multi-time space.  
Each  of these  time evolutions  are associated to a non-linear  equation of motion and henceforth  constitute an integrable hierarchy of equations with common conservation laws.
Another peculiar feature of integrable models is the existence of B\"acklund Transformations  which relate two 
different  field configurations of certain  non-linear differential equation.  

B\"acklund transformations (BT), among other applications, generate an infinite sequence of soliton solutions  
from a non-linear superposition principle (see \cite{rogers}). 
These transformations have also been employed to describe  integrable defects \cite{Corr1} in the sense that two solutions of an integrable model may be interpolated  
 by a defect at certain  spatial position.  
 A BT  connecting a two  field configurations is the key ingredient  to preserve  the integrability of the system. Therefore, its systematic construction
 is important for the classification of integrable defects.

 The first type of B\"acklund transformation only involves the fields of the bulk theory, and is named type I.
 However, there exist integrable models for which such type of B\"acklund transformation are not allowed.  This is the case of the Tzitzeica  model where additional 
 auxiliary fields are required \cite{Corr1, Corr2, pavlov,Corr3,Leandro}.  These   are  called type II and consist of a new class of B\"acklund transformations.
 For the sine(sinh)-Gordon  model, where type I B\"acklund transformation exists, the type II B\"acklund transformation  is 
 shown to be constructed  from the composition of two type I defects. 
 
The novelty  presented in  this paper  is  to  extend the composition  of several  consecutive Gauge-Bäcklund 
transformations for the sinh-Gordon model.  This provides  the  generalization  to the case of $n$ defects by constructing the 
general defect matrix, as well as the corresponding general BT.  This is a powerful method since  
 the defect matrix  appears to be universal  and can be  used  as a generator of BT for all equations within a hierarchy \cite{retore}. Finally, we will present few  solutions for such composite BT.

\section{Gauge-B\"acklund Transformation and Defect Matrices}
The Lax pair for the sinh-Gordon model is given  by
\begin{eqnarray}\label{Lax}
	A_+(\ph_i)=\left(\begin{array}{cc}\pp_+\ph_i & 1\\\\
		\lambda & -\pp_+\ph_i
	\end{array}\right), \qquad A_-(\ph_i)=\left(\begin{array}{cc}0 & \frac{e^{-2\ph_i}}{\lambda}\\\\
	e^{2\ph_i} & 0
\end{array}\right).
\end{eqnarray}
where  we denote $\phi_0$ and $\phi_1$ to be   solutions  for  $x<0$ and $x>0$ regions, respectively.
The defect is placed at $x=0$  and connects  the two solutions by B\"acklund transformation.  We  assume  the Lax pairs  to be related by gauge transformation, i.e.,
\begin{eqnarray}
  K(\phi_0, \phi_1)A_{\pm}(\phi_1) =  A_{\pm}(\phi_0)K(\phi_0, \phi_1) + \pp_{\pm}K(\phi_0, \phi_1) \label{gauge}
\end{eqnarray}
 where  defect matrix describing the transition from  solutions $\phi_0$   to $\phi_1$ is given by
\begin{eqnarray}
K_i \equiv   K(\phi_{i-1}, \phi_i) = \left(\begin{array}{cc}
1 & -\frac{\s_i}{\lambda}e^{-(\phi_{i-1}+\phi_{i})}\\
-\s_i e^{(\phi_{i-1}+\phi_i)} & 1
\end{array}\right).
\label{k1}
\end{eqnarray}
\noindent
and  $\s_i$ is the corresponding B\"acklund parameter. 
 The gauge transformation  (\ref{gauge}) holds  provided  the following first order equations are satisfied, 
\begin{equation}
	\pp_+(\ph_0-\ph_1)=-2\s_1\sinh (\ph_0+\ph_1), \quad \text{and}\quad
	\pp_-(\ph_0+\ph_1)=-\frac{2}{\s_1}\sinh (\ph_0-\ph_1),
	\label{tipo1}
\end{equation}
\noindent
where $ \partial_\pm=\frac{1}{2}(\partial_x\pm\partial_t) $.
Equations \eqref{tipo1} are the type I B\"acklund transformations  for the sinh-Gordon model.

Let us now consider the composition of two  B\"acklund-gauge transformations $ K^{(2)}(\phi_0, \phi_2)= K(\phi_1, \phi_2)\, K(\phi_0, \phi_1)$. From expression (\ref{k1})  we find
\begin{eqnarray}
 	K^{(2)}
 =\left(
 \begin{array}{cc}
 	1+\frac{\sigma _1 \sigma _2 }{\lambda }\, e^{p_1-p_2}  & -\frac{1}{\lambda}\left(\sigma _1 e^{-p_1} +\sigma _2 e^{-p_2} \right) \\\\
 	-\sigma _1 e^{p_1} -\sigma _2 e^{p_2}  & 1+\frac{\sigma _1 \sigma _2}{\lambda }\,  e^{-p_1+p_2} 
 \end{array}
 \right).
 \label{KK}
\end{eqnarray}
Denoting $\eta = {{{\sigma_1^2 + \s_2^2}\over {\sigma_1\sigma_2}}}$,  $\sigma^2 = -{{1}\over {\s_1\s_2}}$ and defining
\br
\Lambda =-\phi_1 - \ln \( 2\s_2 e^{-\phi_0} + 2 \s_1 e^{-\phi_2}\) - \ln {{\s}\over {4}},
\er
we obtain the type II Bäcklund transformations proposed in \cite{Corr3,Thiago}, namely,
\br
  K^{(2)}(p,q,\Lambda)  =  \left( \begin{array}{cc}  1- {{1}\over {\sigma^2 \lambda}}e^q & \quad {{e^{\Lambda -p}}\over {2\lambda\s}}(e^q + e^{-q} + \eta ) \\ \\
  -{{2}\over {\s}} e^{p-\Lambda} & \quad 1- {{1}\over {\lambda \s^2}}e^{-q} \end{array} \right),
 \label{type2mkdva}
 \er
where $q=\phi_0-\phi_2, \quad p=\phi_0+\phi_2$.

Now we consider a system with $n$ Type-I defects, each one with a different parameter $\s_i$ as showing in the following the diagram,
\begin{figure}[h]
\begin{center}
\includegraphics[width=10cm]{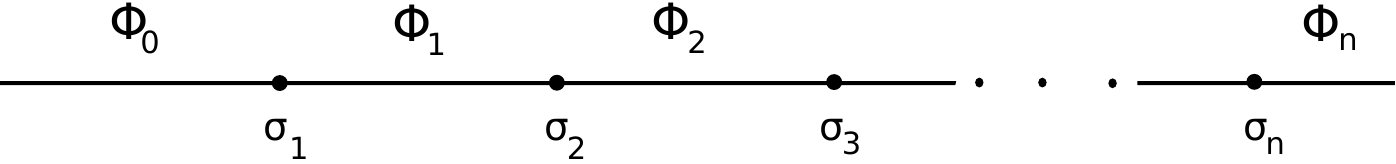}
\caption{Generalization for $n$ type-I defects.}
\end{center}
\end{figure}

\noindent By defining $K^{(n)}$ in the following form,
\begin{eqnarray}
K^{(n)} =K_nK_{n-1}...K_2K_1= \left(
\begin{array}{cc}
K_{11}^{(n)} & K_{12}^{(n)}  \\\\
K_{21}^{(n)} & K_{22}^{(n)}
\end{array}
\right),
\end{eqnarray}
we find for even $n$:
\begin{eqnarray}
K_{11}&=&1+\left[\prod_{a=1}^{n_{\s}}\left(\sum_{i_a=a}^{n-(n_{\s}-a)}\right)\right]\left[\sum_{r=1}^{n/2}\frac{1}{\lambda^r}\prod_{j=1}^{2r}\s_{i_j}e^{(-1)^{j+1}p_{i_j}}\right], \non\\
K_{12}&=&-\left[\prod_{a=1}^{n_{\s}}\left(\sum_{i_a=a}^{n-(n_{\s}-a)}\right)\right]\left[\sum_{r=0}^{(n-2)/2}\frac{1}{\lambda^{r+1}}\prod_{j=1}^{2r+1}\s_{i_j}e^{(-1)^{j}p_{i_j}}\right], \non\\
K_{21}&=&-\left[\prod_{a=1}^{n_{\s}}\left(\sum_{i_a=a}^{n-(n_{\s}-a)}\right)\right]\left[\sum_{r=0}^{(n-2)/2}\frac{1}{\lambda^{r}}\prod_{j=1}^{2r+1}\s_{i_j}e^{(-1)^{j+1}p_{i_j}}\right], \non\\
K_{22}&=&1+\left[\prod_{a=1}^{n_{\s}}\left(\sum_{i_a=a}^{n-(n_{\s}-a)}\right)\right]\left[\sum_{r=1}^{n/2}\frac{1}{\lambda^r}\prod_{j=1}^{2r}\s_{i_j}e^{(-1)^{j}p_{i_j}}\right] ,
\end{eqnarray}    
and for odd $n$:
\begin{eqnarray}
K_{11}&=&1+\left[\prod_{a=1}^{n_{\s}}\left(\sum_{i_a=a}^{n-(n_{\s}-a)}\right)\right]\left[\sum_{r=1}^{(n-1)/2}\frac{1}{\lambda^r}\prod_{j=1}^{2r}\s_{i_j}e^{(-1)^{j+1}p_{i_j}}\right] ,\non\\
K_{12}&=&-\left[\prod_{a=1}^{n_{\s}}\left(\sum_{i_a=a}^{n-(n_{\s}-a)}\right)\right]\left[\sum_{r=0}^{(n-1)/2}\frac{1}{\lambda^{r+1}}\prod_{j=1}^{2r+1}\s_{i_j}e^{(-1)^{j}p_{i_j}}\right] ,\non\\
K_{21}&=&-\left[\prod_{a=1}^{n_{\s}}\left(\sum_{i_a=a}^{n-(n_{\s}-a)}\right)\right]\left[\sum_{r=0}^{(n-1)/2}\frac{1}{\lambda^{r}}\prod_{j=1}^{2r+1}\s_{i_j}e^{(-1)^{j+1}p_{i_j}}\right], \non\\
K_{22}&=&1+\left[\prod_{a=1}^{n_{\s}}\left(\sum_{i_a=a}^{n-(n_{\s}-a)}\right)\right]\left[\sum_{r=1}^{(n-1)/2}\frac{1}{\lambda^r}\prod_{j=1}^{2r}\s_{i_j}e^{(-1)^{j}p_{i_j}}\right] ,
\end{eqnarray}    
where $ n_{\s} $ is the number of parameters $\s_{i_j}$ associated with 
each defect such that $ i_1<i_2<i_3<...<i_n $, and  $p_{i_j}=\ph_{i_j}+\ph_{i_j-1}$.


The next step is to derive a general expression for the BT corresponding to this $K^{(n)}$ defect matrix.
  In order to obtain the Bäcklund transformations for $n$ defects, we will  assume  $K^{(n)}$ to be the generator of  the gauge transformation (\ref{gauge}), 
  leading to 
  \begin{eqnarray}
  	\pp_+(\ph_0-\ph_n)&=&-2\sum_{i=1}^{n}\s_i\sinh p_i\non\\
  	\pp_-(\ph_0-(-1)^n\ph_n)&=& 2\sum_{i=1}^{n}\frac{(-1)^n}{\s_i}\sinh q_i\non\\
  	\pp_+q_i&=&-2\s_i\sinh p_i\non\\
  	\pp_-p_i&=&-\frac{2}{\s_i}\sinh q_i \label{ndef}
  \end{eqnarray}
  with $p_i=\ph_{i-1}+\ph_i, \,\,\, q_i=\ph_{i-1}-\ph_i$,  and $i=1,...,n$.\
  

  \section{Bäcklund solutions}
  In this section we will consider some solutions of the sinh-Gordon model in the presence of two and three defects. 
  
    
 {$\bf {n=2}$}  
 
 Consider now the fields $\ph_0$ and $\ph_2$ on each side of the defect with   an intermediary field $\ph_1$,

  {\bf{Vacuum $ \rightarrow $ One Soliton $ \rightarrow $ Vacuum  Solution.}}  First of all, we consider the following solution:
  \begin{eqnarray}
  \ph_0= 0,\quad\ph_2=0,\quad \ph_1=\ln\left(\frac{1+\rho_1}{1-\rho_1}\right),\quad \rho_1=\exp \left(2k_1 x_{+}+\frac{2}{k_1}x_{-}\right),
  \end{eqnarray}
which satisfy  the Bäcklund equations (\ref{ndef}) with  $n=2$ with the following conditions
  \begin{eqnarray}
  \s_1=k_1,\quad \s_2=-k_1.
  \end{eqnarray}

{\bf{Vacuum $ \rightarrow $ One Soliton$ \rightarrow $ Two Soliton Solution.}}
Another  possible solution is
  \begin{eqnarray}
  \ph_0= 0,\quad\ph_1&=& \ln\left(\frac{1+\rho_1}{1-\rho_1}\right),\quad \ph_2=\ln\left(\frac{1+b_1 \rho_1+b_2 \rho_2+\a_{12} b_1 b_2 \rho_1 \rho_2}{1-b_1 \rho_1-b_2 \rho_2+\a_{12} b_1 b_2 \rho_1 \rho_2}\right),\non\\
  \rho_j&=&\exp \left(2k_j x_{+}+\frac{2}{k_j}x_{-}\right),\quad j=1,2
  \end{eqnarray}
  where, in order to $\ph_2$ satisfies the sinh-Gordon equation $ \a_{12}=\left(\frac{k_1-k_2}{k_1+k_2}\right)^2. $
    Analogously, we get the following  Bäcklund conditions:
  \begin{eqnarray}
  \s_1=k_1,\quad\s_2=k_2,\quad b_1=\frac{k_1+k_2}{k_1-k_2}.
  \end{eqnarray}

{$\bf {n=3}$}  

Finally putting a third defect at the same point of the others we have the fields $\ph_0$ and $\ph_3$ on each side of the defects and two intermediary fields $\ph_1$ and $\ph_2$ at the defect points.  
  
   {\bf{Vacuum $ \rightarrow $ One Soliton$ \rightarrow $ Vacuum $ \rightarrow $ One Soliton Solution.}}
  Now taking into account the solutions:
  \begin{eqnarray}
  	\ph_0=0,\quad \ph_2=0,\quad \ph_1=\ln\left(\frac{1+\rho_1}{1-\rho_1}\right),\quad\ph_3=\ln\left(\frac{1+\rho_2}{1-\rho_2}\right)
  \end{eqnarray}
  The Bäcklund conditions in order to satisfy the Type-II BT are: $\s_1=k_1,\, \s_2=-k_1,\, \s_3=k_2$.
  
     {\bf{Vacuum $ \rightarrow $ One Soliton$ \rightarrow $ Two Soliton $ \rightarrow $ Three Soliton Solution.}} Lastly, we assume:
  \begin{eqnarray}
  	\ph_0 &=& 0,\quad\ph_1= \ln\left(\frac{1+\rho_1}{1-\rho_1}\right),\quad \ph_2=\ln\left(\frac{1+b_1 \rho_1+b_2 \rho_2+\a_{12} b_1 b_2 \rho_1 \rho_2}{1-b_1 \rho_1-b_2 \rho_2+\a_{12} b_1 b_2 \rho_1 \rho_2}\right),\non\\
  	\ph_3&=&\ln\left(\frac{1+R_1+R_2+R_3+\a_{12} R_1 R_2+\a_{13} R_1 R_3 +\a_{23} R_2 R_3 +\a_{123} R_1 R_2 R_3}{1-R_1-R_2-R_3+\a_{12} R_1 R_2 +\a_{13} R_1 R_3 +\a_{23} R_2 R_3 -\a_{123} R_1 R_2 R_3}\right),\non\\
  	\rho_j&=&\exp \left(2k_j x_{+}+\frac{2}{k_j}x_{-}\right),\quad R_j=a_j\rho_j, \quad j=1,2,3
  \end{eqnarray}
  where in order to $\ph_2$ and $\ph_3$ satisfy the sinh-Gordon equation we should have 
  \begin{eqnarray}
  	\a_{12}=\left(\frac{k_1-k_2}{k_1+k_2}\right)^2,\, \a_{23}&=&\left(\frac{k_2-k_3}{k_2+k_3}\right)^2,\, \a_{13}=\left(\frac{k_1-k_3}{k_1+k_3}\right)^2,\non\\ \a_{123}&=&\a_{12}\a_{13}\a_{23}.
  \end{eqnarray}  
In this case the Bäcklund conditions are:  $ \s_1=k_1,\quad\s_2=k_2,\quad\s_3=k_3,\quad b_1=\frac{k_1+k_2}{k_1-k_2}, $ $   a_1=\left(\frac{k_1+k_3}{k_1-k_3}\right)b_1$, and $a_2=\left(\frac{k_2+k_3}{k_2-k_3}\right)b_2$,
\noindent
where $ b_2 $ is a free parameter. 
It is worth mentioning that the BT and their solutions for a four-defect system have been also computed, and the results have shown the expected behaviour. 

\section{Conclusion}

In this paper we have  considered the sinh-Gordon model  and provided general formulas for the defect matrix  when  n defects are considered.
Our construction involves the  product of n Type-I defect matrices.  In addition, we have calculated their respective BT in a general way through gauge transformations and 
provided a few simple examples for $n=2,3$.

It is important to point out that, since  the BT  are constructed  as gauge transformations,
they preserve the zero curvature representation. The later describes  a hierarchy of integrable  equations  based upon an universal   Lax operator.  
These two facts induces  the idea of the universality of the B\"acklund-Gauge transformation within the hierarchy.  
We have verified \cite{Ana-1, Ana} that the constructed   defect matrix indeed  gives the correct  BT for the mKdV equation. It  provides a systematic  construction of BT for all   higher grade evolution equations within the mKdV hierarchy.
Several examples were verified   for  KdV hierarchies \cite{retore} as well.


\begin{acknowledgement}

The authors would like to thank the organizers of the colloquium ICGTMP - Group 31 for the opportunity to present our work. ALR would like to thank FAPESP São Paulo Research Foundation for the financial support under the process 2015/00025-9. JFG would like to thank FAPESP and CNPq for the financial support. NIS and AHZ would like to thank CNPq for the financial support. 

\end{acknowledgement}

\end{document}